\documentstyle[11pt,aaspp4]{article}

\def\ha	{H$\alpha$}

\def\sii	{[S~{\sc ii]}}
\def\nii	{[N~{\sc ii]}}

\def\oiii	{[O~{\sc iii]}}

\def\kms	{km~s$^{-1}$}

\def\sb  {ergs cm$^{-2}$ s$^{-1}$ arcsec$^{-2}$}

\def\per#1{$^{-#1}$}

\def\etal	{et~al.}

\def \amin	{\rlap{$.$}'}
\def \asec	{\rlap{$.$}''}

\def \hst    {{\it HST}}

\def \ein	{{\it Einstein Observatory}}

\lefthead{Smith}
\righthead{RCW 86}

\begin{document}

\title{The Discovery of Balmer-filaments Encircling SNR 
RCW 86}

\author{R. Chris Smith\altaffilmark{1,2}}
\affil{Department of Astronomy, University of Michigan \\
Ann Arbor, MI 48109-1090\\
chris@astro.lsa.umich.edu}

\altaffiltext{1}{McLaughlin Fellow, University of Michigan.} 

\altaffiltext{2}{Also at Cerro Tololo Inter-American Observatory 
(CTIO).  CTIO is a division of the National Optical Astronomical 
Observatories (NOAO), operated by AURA, Inc.\ under contract to the 
National Science Foundation.}

\bigskip
\bigskip
\begin{center}
Accepted for publication in \\
{\it The Astronomical Journal}\\
\end{center}

\begin{abstract}

We report the discovery of Balmer-dominated filaments along almost the 
complete periphery of the supernova remnant RCW~86 (also known as 
G~315.2-2.3 or MSH~14-6{\it 3}).  Using the UM/CTIO Curtis Schmidt 
telescope, we obtained deep CCD images in the emission of \ha\ and 
\sii, together with continuum images to remove stellar confusion.  
After continuum subtraction, we discovered a network of \ha\ filaments 
reaching almost completely around the remnant.  Most of the newly 
identified filaments show no corresponding \sii\ emission, indicating 
that they belong to the peculiar ``Balmer-dominated'' class of 
filaments.  Comparison of these Balmer filaments with existing radio 
and X-ray images of RCW~86 shows an overall similarity, although 
interesting differences are apparent upon detailed inspection.  While 
further observations of these newly identified optical filaments may 
eventually provide more detailed information on the kinematics and 
distance to RCW~86, the distance currently remains uncertain.  We 
argue that the shorter distance estimates of $\sim$1 kpc are still 
favored.

\end{abstract}

\keywords{supernova remnants: individual (RCW~86) --- ISM: individual
(RCW~86) --- nebulae: individual (RCW~86) --- shock waves ---
supernovae: individual (SN 185)}

\vfill\eject

\section{Introduction}


The supernova remnant RCW~86 (also known as MSH~14-6{\it 3}, 
G~315.4-2.3, PKS 1439-62) has received much attention in recent years.  
It is among the brightest supernova remnants (SNRs) in X-rays in our 
Galaxy, and shows up as a bright non-thermal shell in the radio.  
Perhaps more interesting, though, is that it has been identified as 
the the possible remnant of the supernova of A.D.\ 185, the first 
historical report of a Galactic supernova (\cite{cs77}).  However, the 
relation between the supernova and this remnant has been called into 
question recently (e.g., \cite{hm-s87} and \cite{thorsett92}), and 
indeed the nature of the recorded event has also been questioned 
(\cite{ch94}, \cite{schaefer95}).  Much of the debate revolves around 
the distance to the supernova remnant, with values which range from 
$<$1 to 3 kpc, and whether the event of A.D.\ 185 was a Type Ia or 
Type II supernova, if it was in fact a supernova!  But the connection 
between the observations of a bright object in A.D.\ 185 and the 
SNR RCW~86 remains an intriguing possibility which, if confirmed, 
would provide an exact age for the remnant, making it the 
oldest SNR with a known age and allowing for more detailed modeling 
and a better understanding of its evolution.


The SNR RCW~86 is a bright, almost complete shell in 
both radio and X-rays.  The radio shell was identified in the MSH 
survey (\cite{msh61}) and soon after found to be a non-thermal and 
polarized source (\cite{hill64}, 1967\nocite{hill67}), indicating that 
it was in fact a supernova remnant.  Later higher resolution radio 
work (e.g., Kesteven \& Caswell 1987) showed that the remnant is a 
roughly circular shell, approximately 40$'$ in diameter, with a bright 
knot in the southwest.  Distance estimates based on the radio $\Sigma$ 
-D relation (e.g., \cite{cc76}, \cite{il72}, \cite{milne70}) gave 
2.0--3.2 kpc.  While the $\Sigma$ -D relation has been shown to be 
notoriously unreliable (\cite{green91}), many of the distances which 
continue to be quoted in the SNR literature are based on these 
estimates.


The first X-ray observations of RCW~86 were obtained by Naranan et al.  
(1977\nocite{naranan+77}), who interpreted the 0.5-2.5 keV X-ray 
spectrum as thermal bremsstrahlung radiation with a temperature of 
$\sim$0.4 keV. The \ein\ provided the first spatially resolved X-ray 
maps of the remnant (\cite{pisarski+84}).  The IPC observations showed 
a limb-brightened shell in general agreement with the overall radio 
structure, though significant detailed differences were apparent.  
Both IPC and HRI observations of the southwest portion of the remnant 
showed a bright knot (or knee, as resolved in the HRI observation), 
roughly matching the bright radio emission from that region.  Analysis 
of the \ein\ dataset and subsequent X-ray observations 
(\cite{claas+89}, \cite{kaastra+92}) have all agreed on distances of 
between 0.7 and 1.3 kpc {\it assuming} an age of $\sim$1800 years, 
i.e., that the remnant is that of SN 185.


While the remnant is a roughly circular shell with a bright knee in 
the SW in both radio and X-rays, the previously known optical emission 
was confined to a complex of bright filaments in the southwest, a 
couple of knots of optical emission to the west of the center, and a 
few additional short filaments along the northern edge of the 
radio/X-ray remnant.  Actually, the original designation ``RCW~86'' 
referred specifically to the bright optical nebulosity identified by 
Rodgers, Campbell, \& Whiteoak (1960\nocite{rcw60}) in the southwest 
of the remnant, filaments which correspond to the brightest radio and 
X-ray structures.  However, herein we use the designation RCW~86 to 
refer to the whole SNR, as is now common in the literature.  
Although the relation of the two knots and northern filaments to the 
remnant was noted as long ago as 1967 (\cite{hill67}), almost all 
optical studies of the remnant have concentrated on only the bright 
filaments in the southwest.  Westerlund \& Mathewson 
(1966\nocite{wm66}) established that these optical filaments showed 
strong \sii\ $\lambda$6724 emission, a common characteristic of 
SNR filaments.  Later spectroscopy (\cite{ruiz81}, \cite{ld83}) 
confirmed that the emission from the southwestern nebulosity was 
consistent with that calculated from models of radiative shock 
emission.


More recently, Long \& Blair (1990\nocite{lb90}) discovered a
different type of filaments, so-called Balmer-dominated filaments, in
the southern edge of the bright southwestern complex and also among
the northern filaments.  Balmer-dominated filaments are distinguished
by a spectrum of Balmer lines with little or none of the
forbidden-line emission (such as \sii, \nii, and \oiii) that
characterizes the radiative filaments typical of most supernova
remnants.  Such Balmer-dominated filaments arise from relatively high
velocity shocks passing through low-density, partially neutral gas,
producing a collisionless shock front.  Neutral atoms (mostly
hydrogen) stream through the shock front into the hot postshock
medium, where they are almost immediately ionized.  However, there is
a small probability that the atoms are excited and emit before being
ionized, producing the faint Balmer filaments seen in a small minority
of remnants.  Such filaments define the Balmer-dominated class of
SNRs, which includes SN 1006, Tycho, and four remnants in the LMC
(\cite{smith+91} and references therein).  Balmer-dominated filaments
have also been observed in Kepler's SNR (\cite{fesen+89}) and along
the outer edges of the Cygnus Loop (\cite{raymond+83}, \cite{hrd86}).
While the radiative filaments typical of evolved SNRs represent gas in
the cooling zones behind the shock front, these Balmer-dominated
filaments are produced at the leading edge of the shock, where the gas
is in the process of being ionized.  Hence these Balmer filaments
trace the actual shock front as it moves into the ambient interstellar
medium.

Here we report the discovery of a complex network of Balmer-dominated 
filaments stretching around almost the complete periphery of RCW~86.  
These newly identified filaments provide a detailed, high-resolution 
optical tracer of the blast-wave shock front, and offer opportunities 
to better understand the remnant and its possible progenitor.  In \S 
2, we present the observations and reductions, and in \S 3 we go on to 
describe the newly identified optical filaments.  In \S 4 we compare 
the optical outline of the SNR with the morphologies in both the radio 
and X-rays.  A summary is given \S 5, together with a review of the 
current debate over the distance to and possible progenitor of SNR 
RCW~86.

\section{Observations \& Reductions}


The narrow-band optical images of RCW~86 were taken with the CCD
camera attached to the Newtonian focus of the 0.6/0.9m UM/CTIO Curtis
Schmidt telescope during two separate observing runs, 1994 March 23-24
and 1995 January 26-29 UT. For both runs, the detector was a Thomson
1024$\times$1024 CCD with 19$\micron$ pixels, providing a scale of
1$\asec$835 pixel\per1 and a field of view of 31$\amin$3.  The
resulting FWHM resolution for the observations averaged $\sim$1.8
pixels or $\sim$3''.  Narrow bandpass filters of \ha\ ($\lambda_{c}$ =
6563 \AA, FWHM = 26 \AA) and \sii\ ($\lambda_{c}$ = 6718 \AA, FWHM =
50 \AA) were used to isolate the nebular emission.  In addition, a
``red continuum'' filter ($\lambda_{c}$ = 6844 \AA, FWHM = 93 \AA )
was used to obtain images free from emission lines, to allow for
accurate subtraction of stars from the emission-line images.

Since the field of view was smaller than the $\sim$40$'$ diameter of 
RCW~86, a rough grid of four fields was used to cover the area of the 
remnant, with one additional field near the center for improved 
signal-to-noise in the center and more accurate image alignment.  
Multiple images with small offsets were taken through each filter at 
each pointing.  A summary of the observations is provided in Table 
\ref{obssum}.  Images of several spectrophotometric standards 
(\cite{hamuy+92}) were obtained to provide absolute flux calibration 
for the emission-line images.


The images were reduced using standard IRAF\footnote{IRAF is
distributed by the National Optical Astronomy Observatories, operated
by AURA, Inc., under contract from the National Science Foundation}
tasks for zero correction, bias subtraction, and flat-fielding with
twilight sky flats.  The images were then combined into mosaics using
IRAF's {\tt geotrans} routine and a procedure similar to that
described by Winkler, Olinger, \& Westerbeke
(1993\nocite{wow93}). This involved identifying several ($\sim$10)
stars with known positions (from the \hst\ Guide Star Catalog) in each
image and then mapping the images individually into ``canvases'' the
size of the final image.  These canvases were then combined to produce
the final \ha, \sii\, and red continuum mosaics.  The resulting \ha\
mosaic image is shown in Fig.\ \ref{hawstars} (Plate XX).


To remove the stellar confusion, $\sim$100 stars from the mosaic of 
images taken in the red continuum filter were measured and compared to 
the stars in the \ha\ and \sii\ mosaics.  The red mosaic was then 
scaled by the resulting average flux ratio and subtracted from each 
emission-line mosaic.  The resulting continuum-subtracted \ha\ mosaic is 
presented in Fig.  \ref{hacsub} (Plate XX).  To help remove the 
remains of subtracted stars, the emission-line mosaics were median 
filtered with a 5$\times$5 pixel (9$''$ square) kernel.  The final 
median-filtered \ha\ and \sii\ mosaics are shown in Figures 
\ref{hamsub} (Plate XX) and \ref{siimsub} (Plate XX), respectively.  
The \ha\ mosaic has an average per pixel exposure time of $\sim$5400s 
(not counting the center, where the exposure is much deeper due to the 
overlap of the fields), while the average \sii\ exposure time per 
pixel is $\sim$1900s.  In the continuum-subtracted \ha\ image, which 
is the image we use to measure surface brightnesses, the background 
rms noise is approximately $1.2\times10^{-17}$ \sb\ (not including bad 
stellar subtractions).  This pixel-to-pixel noise, however, provides 
only a conservative surface brightness limit estimate for our \ha\ 
image, since large-scale features covering many pixels even at that 
limit will be easily visible.

\section{Optical Morphology of RCW~86}

The final continuum-subtracted \ha\ mosaic image is reproduced in Fig.
\ref{hawlab} with the principal filaments and filament complexes
labeled.  The filament complexes have been numbered from the geometric
center of the remnant outward, so that smaller numbers indicate
smaller projected distance from the center, and a suffix ``b'' or
``r'' has been added to indicate Balmer-dominated and radiative
filaments, respectively.  Where there is a mix of radiative and
Balmer-dominated filaments, we leave off the suffix unless we are
speaking specifically of one or the other.  In lieu of spectroscopic
information, the filaments which show no \sii\ counterparts in Fig.\
\ref{siimsub} are considered Balmer-dominated.  While the \sii\ mosaic
is not as deep as the \ha\ mosaic, our \sii\ images are nevertheless
sensitive enough to identify filaments with a \sii/\ha\ ratio of
$\ge$0.4, characteristic of radiative shock emission, for all of the
\ha\ filaments discussed herein.


It is immediately apparent that these newly discovered
Balmer-dominated filaments, stretching almost all the way around the
periphery of RCW~86, provide a much more complete picture of the
remnant than do the radiative filaments.  Indeed, even at the
relatively coarse resolution of 3$''$, the \ha\ mosaic provides the
highest resolution image of the complete shell of RCW~86 available at
any wavelength to date.  The west and east sides are well defined,
giving the remnant a somewhat boxy appearance, with only the
southeastern corner showing little detectable optical emission.

As shown by Long \& Blair (1990\nocite{lb90}), Balmer-dominated
filaments are intermingled with radiative filaments in the edges of
SW1 (the original RCW~86 nebula).  The previously identified filaments
are some of the brightest Balmer-dominated filaments of the whole
remnant, with surface brightnesses of up to $\sim3.0\times10^{-16}$
\sb.  Additional fainter Balmer filaments (S1b) are seen extending to
the east of the bright nebula, tracing out at least part of the
southern portion of the remnant.


Along the northern side, we see that the filaments break up into two 
separate, almost parallel, complexes.  Long \& Blair identified the 
Balmer filaments mixed in with the radiative filaments along the inner 
set of northern filaments, N1.  To the north of this mixture are 
several knots of emission, which we label N2.  Given the apparent 
leading position of this outer set of filaments, there are 
surprisingly few Balmer filaments detected among the radiative 
filaments.  Some of the filaments seem to be Balmer-dominated at one 
end and radiative at the other, perhaps implying a recent interaction 
of the remnant shock front with dense clumps of matter.


The filaments of the N1 complex can be followed down the east side 
along a very faint, somewhat diffuse filament which brightens and 
sharpens into E2b, the outer of the eastern filaments.  The E2b 
complex is actually two or three parallel filaments with faint Balmer 
emission filling the area between them.  To the interior is the 
fainter E1b filament.  Since Balmer-dominated filaments trace the 
blast wave shock front, where neutral gas is being ionized, E1b's 
interior position is most likely only the result of projection.  It 
probably represents a ripple along the front or back face of the 
remnant shell.  A similar argument might be made for the inner 
filament along the northern rim, N1, but the broken structure and lack 
of many Balmer filaments in N2 implies that the situation is probably 
more complex there.


The peculiar emission-line profile which characterizes
Balmer-dominated filaments may provide a key to deciphering the
three-dimensional geometry of these shocks.  The \ha\ line profile
from these collisionless shocks consists of two components
(\cite{cr78}, \cite{ckr80}, \cite{kwc87}, \cite{smith+91}).  A narrow
component is produced as neutral H atoms stream through the shock
front and are excited in the hot post-shock medium {\it before} they
are ionized and accelerated to the post-shock velocity distribution.
The line profile from these ``slow neutrals'' is therefore
characteristic of the pre-shock conditions, and typically has a FWHM
of $\sim$30 \kms\ (\cite{slr94}) and a line center corresponding to
the rest velocity of the pre-shock gas.  The second component is a
broad feature created by fast ions which undergo charge exchange with
the slow neutrals, producing ``fast neutrals'' which have a small
probability of being excited and emitting before being ionized.  This
broad component has a FWHM proportional to the shock velocity and a
line center which is determined by the combination of the bulk flow of
the post-shock gas and the angle to the line of sight.  The offset
between the narrow-component line center and that of the broad
component is therefore the component of the post-shock flow along our
line of sight.  Given that we can determine the velocity of the
post-shock flow from the width of the broad component, we can derive
the angle to the line of sight for any filament which shows an offset
between the line centers of the broad and narrow components.  The
offset measured by Long \& Blair (1990\nocite{lb90}) of 1.4 \AA\ in
the N1b filaments translates to an angle to our line of sight of
approximately 84$^{\circ}$.  Assuming an approximately spherical
shell, this places those filaments in a position on back face of the
shell about 6$^{\circ}$ behind the plane of the sky.  Similar
measurements should allow us to determine the relative positions of the
E1b and E2b filaments.


The west side of the remnant exhibits a far more complex structure, 
with three distinct filament complexes.  The W3b filament complex 
seems to mark the projected outer extent of the remnant, where the 
shock is presumably close to perpendicular to our line of sight.  It 
is made up of many short filaments, a few of which are at large angles 
to the average curve, creating a small triangular structure in the 
west-northwest.  The W2b filament complex, the brightest of the three 
structures, is much more uniform, consisting of several long, thin 
filaments.  These are again presumably ripples either in the front or 
back shell of the remnant.

The whole area of the western filaments is filled with a faint diffuse
component, with the W1b filaments marking the inner edge of this
diffuse emission.  If we assume that this emission is from the uniform
surface of the shell between W2b and W1b, the \ha\ intensity provides
a first order method to estimate the neutral ambient density,
$n_{HI}$.  Using the relation given in Winkler \& Long
(1997\nocite{wl97}, eq.  2), which is based upon the Chevalier \&
Raymond (1978\nocite{cr78}) derivation of 0.048 \ha\ photons per H
atom passing through the shock, we find that $$ I_{\perp} = 0.23\times
10^{-16} n_{HI} (v_{s}/800 {\rm\ km\ s}^{-1}) {\rm\ ergs\ cm}^{-2}
{\rm\ s}^{-1} {\rm\ arcsec}^{-2}.$$ Measuring the surface brightness
midway between W1b and W2b, at a projected radius of approximately
$R/R_{SNR}=0.7$, the intensity should be approximately $I = 3\times
I_{\perp}$ (assuming a line of sight passing through a thin spherical
shell, $\Delta R/R < 0.01$).  The measured surface brightness of
$6\times 10^{-17}$ \sb, together with the shock velocity of 800 \kms\
measured in N1 (from \cite{lb90}) gives us an estimate of $n_{HI}
\approx 0.9$ cm\per3, which is slightly higher than than found in
either Tycho's SNR (\cite{kwc87}) or SN 1006 (\cite{wl97}).  This is
of course only a rough estimate.  With detailed spectroscopy,
resulting in measurements of the broad and narrow component line
centers and an estimate of the shock velocity at the same location, we
should be able to determine the geometry of these filaments, map out
the 3-dimensional structure of the remnant shell in this region, and
derive better limits on the ambient gas density and its variation.

Mixed in with all of the Balmer-dominated filaments are two knots of 
radiative shock emission, labeled W1r.  These knots were noted by 
Hill (labeled ``B'' in \cite{hill67}), but as with the northern 
radiative filaments, no spectroscopic observations of them have ever 
been published.  


Finally, there are many extremely low level, diffuse structures within
the boundaries of RCW~86.  These have \ha\ surface brightnesses of $\sim3 -
6\times10^{-17}$ \sb, still well above the surface brightness limit of
the mosaic ($\sim1\times10^{-17}$ \sb).  Perhaps the most
interesting of these is a large triangular structure near the
center of the remnant.  This could simply be background (or
foreground) diffuse \ha\ emission, or it could be the front or back
side of RCW~86.  If it is the latter, deep spectroscopy could provide
yet more 3D information about the remnant.  A similar structure shows
up at very low level in the \sii\ image, implying that this faint
central emission is not Balmer-dominated emission.

\section{Radio and X-ray comparison}

The optical shell of RCW~86, as defined by the newly discovered
Balmer-dominated filaments, provides a more complete basis for
comparison with the radio and X-ray images.  While detailed
comparisons await our upcoming high-resolution observations at radio
(with the ATCA) and X-ray (pending ROSAT HRI observations)
wavelengths, we have obtained existing radio and X-ray images in order
to present a preliminary comparison here.  In Fig.\ \ref{3band} we
show a ROSAT PSPC image (observation RP500078A02), which has a
resolution of approximately 30$''$, and the 843 MHz radio image of
Kesteven \& Caswell (1987\nocite{kc87}), with a resolution of
$\sim$45$''$,alongside our continuum-subtracted \ha\ image.  The
global structure at the three wavelengths is similar, although several
interesting differences are apparent.


As previously noted, the radiative filaments in the southwest coincide 
in all three bands, although in this comparison one can see that the 
radio shows less of the knee shaped structure seen in both the optical 
and X-ray images.  Radio (\cite{hill67}), X-ray (\cite{pisarski+84}, 
\cite{claas+89}), optical (\cite{ruiz81}, \cite{ld83}, 
\cite{rosado+96}), and infra-red (\cite{gs90}) studies all agree that 
this bright feature has been produced by the SNR shock front 
encountering a dense cloud.  The interaction is thought to be 
relatively recent because while observations indicate that the shock 
has been significantly decelerated in this region (from the $\sim$600 
to 800 \kms\ found in the northern filaments by \cite{lb90} down to 
$\sim$100 \kms\ in the southwest, as measured by \cite{rosado+96}), 
the complex is at roughly the same distance from the center as the 
rest of the shell.


The eastern rim of the remnant also shows overall agreement at all
three wavelengths.  The sharp outer Balmer filaments of E2 seem to
bound the radio and X-ray emission, but given the resolution of these
radio and X-ray images, we are unable to measure any offset.  The
radio shows a definite filament branching off to the interior,
corresponding to the E1 filament in \ha, and a corresponding feature
in the X-ray appears to be present at a low level.  


Moving north and west along the shell, the agreement between the three
images breaks down somewhat along the northern rim.  The X-ray image
shows a distinct break in the northeast, which is accompanied by both
a break in the optical filaments and an optical filament just beyond
the break, suggestive of a mini-blowout at this location.  The radio
image shows no corresponding blowout, but just beyond that position it
splits into a lower filament, which corresponds to the optical N1
complex as well as the X-ray bright rim, and an upper filament.  The
upper portion of radio emission forms a ``northern cap'', which
extends out beyond the N2 filaments seen in the optical.  Some
evidence of the northern cap is also seen in the X-ray image, although
at a very faint level.  The overall structure of the region, including
the broken optical morphology of both the N1 and N2 optical filament
complexes and the extension of the northern cap outside of the
brightest filaments of radio and X-ray (corresponding to N1), is
suggestive of a ``captured blowout'', in which the shock has rushed
out beyond the average expansion radius of the remnant (possibly
defined by the N1 filaments), only to encounter a denser region.  It
is, however, puzzling that the northernmost edge, as defined by the
radio and X-ray, shows no detected optical emission.


The complex western region shows some of the most striking differences
between the three images.  While faint optical filaments and the radio
trace out the outer shock front W3b, there is little
corresponding X-ray emission in the region.  The X-ray emission is
strongest along W2, which shows no radio counterpart, and fills the area
between the W2 and W1 complexes.  A faint radio filament seems to
correspond to the inner filament W1, looping over to connect to W3
across the top of the region between W1 and W2 filled by both X-ray
and Balmer emission.


Perhaps the most striking difference between the three images is the
radio filament stretching across the southern portion of RCW~86.
While both optical and X-ray filaments lie along the southwestern
portion of this arc, they both fade out, leaving only the radio to
trace out what appears to be a feature crossing the southern face of
the remnant.  This does not appear to be the true southern extent of
the remnant, as there is faint radio and X-ray emission outside of this
feature in the southeast.


\section{The Overall Picture, Distance, and Progenitor of RCW~86}

The radio, X-ray and new optical images provide the general impression 
of RCW~86 as a SNR expanding into a very inhomogeneous medium.  To the 
southwest the blast wave has encountered a rather large, dense cloud, 
and to the north it has encountered smaller scale density enhancements 
embedded in a low-density partially neutral medium, as demonstrated by 
the mixture of radiative and Balmer-dominated filaments observed 
there.  On the east and west sides the remnant is expanding into 
rather low-density, partially neutral gas, but the relatively bright 
X-ray filaments on the west side seem to imply some significant
variations in the conditions even in that region.

These variations in the surrounding ISM are perhaps not surprising, 
given the projected location of RCW~86 relatively close to the 
Galactic plane (l=315.4, b=-2.3; the Galactic plane lies to the 
north-northwest of the remnant).  Its actual distance off of the 
Galactic plane is, of course, dependent on the distance to the SNR, 
the value of which has been the subject of much debate in the 
literature.  All recent X-ray analyses derive distances of $\sim$1 
kpc, while most of the other estimates range from 2.3 to 3.2 kpc.  
While many of the X-ray studies have assumed an age of 1800 years 
(corresponding to the assumption that the remnant is the result of SN 
185), Nugent \etal\ (1984\nocite{nugent+84}) and Leahy 
(1996\nocite{leahy96}) do not base their analyses on this assumption 
and nevertheless derive distances of $\sim$0.7 kpc and 1.6 kpc, 
respectively.

The larger distance estimates are based on the $\Sigma$-D relation, 
the possible link between the SNR and an OB association, a kinematic 
distance, or the measured optical extinction.  In the radio, there are 
no published H I absorption distance estimates, so the only estimates 
available are based on the $\Sigma$-D relation.  However, as mentioned 
in the introduction, $\Sigma$-D is extremely unreliable 
(\cite{green91}), with uncertainties which could place RCW~86 
at a distance of anywhere between 1 and 10 kpc (\cite{strom94}).

One of the oldest, and possibly most influential, distance estimates 
is the suggestion of a link between RCW~86 and an OB association found 
along the line of sight.  Westerlund (1969\nocite{westerlund69}) 
proposed this link based upon the projected proximity of several OB 
stars, the rough agreement between the radial velocities of the 
filaments (+1 and +27 \kms) and the stars (between -14 to +20 \kms), 
and the assumption that the remnant was of a Type II supernova.  
However, the majority of recent X-ray analyses find that a Type Ia 
progenitor is favored by the models (\cite{pisarski+84}, 
\cite{claas+89}, \cite{leahy96}; see however \cite{kaastra+92}), 
although a Type II SN cannot be ruled out.  Also, recent Fabry-Perot 
measurements of the bright filaments in the bright southwestern knee 
of RCW~86 by Rosado \etal\ (1996\nocite{rosado+96}) give a V$_{LSR}$ of 
-33.2$\pm$4.5 \kms, seemingly inconsistent with the reported stellar 
velocities (although it is not clear if those velocities were 
corrected to V$_{LSR}$).  Rosado \etal\ continue on to derive a 
kinematic distance for RCW~86 of 2.8$\pm$0.4 kpc, but this estimate is 
again based upon the assumption that the progenitor SN was of Type II 
(as they point out), and therefore neither defines the SN type
nor provides an unambiguous distance.


Large values of optical extinction have also been pointed to in 
support of larger distances.  Leibowitz \& Danziger 
(1983\nocite{ld83}) derived a value of A$_{V}$=1.7 mag from spectra of 
various bright filaments of the southwestern knee.  As they point out, 
this value is consistent with the A$_{V}$ of 2 measured for the OB 
association (\cite{westerlund69}).  However, these values seem at odds 
with the absorption derived from various X-ray measurements, which 
have consistently yielded column densities of approximately 
$1-2\times10^{21}$ cm\per2 or less (\cite{nugent+84}, 
\cite{pisarski+84}, \cite{claas+89}, \cite{kaastra+92}, 
\cite{leahy96}, \cite{vkb97}).  If we take the simplest conversion provided by 
Burnstein \& Heiles (1978\nocite{bh78}, eq.\ 7), we obtain an $E(B-V)$ 
of $\sim$0.15, or an A$_{V}$ of $\sim$0.5.  Indeed, Leibowitz \& 
Danziger's own extinction data show significant scatter, including at 
least one very low value which they attribute to a hole in the 
foreground material.  As Strom (1994\nocite{strom94}) suggests, the 
scatter in the optical extinction may equally well be attributable to 
absorption in or near the filaments, due to the significant amount of 
dust evident (\cite{gs90}) where the optical extinction was measured.  
Optical measurements of the radiative filaments in the north, away 
from the large dense cloud in the southwest, may provide better 
estimates of the true optical extinction.

The large distance estimates, and the corresponding apparent relation
to the OB association, have recently been used as arguments against
the hypothesis that RCW~86 is the remnant of SN 185
(\cite{thorsett92}, \cite{schaefer95}, \cite{rosado+96}).  However,
the smaller distance of $\sim$1 kpc is certainly not ruled out by
existing observations, and may even be favored over the larger
distance estimates, despite the claims to the contrary published since
Strom's (1994\nocite{strom94}) review, which favored the smaller
distance and argued that RCW~86 is the most likely remnant of SN 185.
Finally, while the lack of proper motion in the bright southwestern
filaments reported by Kamper, van den Bergh, \& Westerlund
(1995\nocite{kvw95}) was said to contradict the youth of RCW~86, it
fails to constrain the overall expansion of the remnant.  As pointed
out above, the shock has been slowed by almost an order of magnitude
in this region (as compared to the northern rim), and thus would not
be expected to have the large proper motion characteristic of a young,
nearby remnant.  On the other hand, the proper motion of the
Balmer-dominated filaments should provide an accurate measurement of
the expansion rate, and when combined with a shock velocity determined
from spectroscopy of the same filaments, should provide an accurate
estimate of the distance (e.g., \cite{lbvdb88}, \cite{smith+91}).

Further observations, especially of these newly identified
Balmer-dominated filaments, will certainly shed further light on the
distance to, and perhaps the progenitor of, SNR RCW 86.  Low
resolution spectroscopic data for the northern radiative filaments
could yield independent optical reddening estimates (away from the
apparently dense cloud in the SW), while moderate resolution spectra
can provide shock velocities and angles to the line of sight for most
of the Balmer-filaments.  High-resolution spectra of the Balmer
filaments would also give a more precise measurement of the systemic
velocity than that of Rosado \etal\ (1996 \nocite{rosado+96}), since
the line center of the narrow component provides an accurate
measurement of the preshock gas conditions independent of the angle to
our line of sight.  And in the near future, we should be able measure
the proper motion of the Balmer-dominated filaments, finally giving an
accurate distance to the SNR.  When combined with detailed X-ray
spectroscopy (e.g., \cite{hughes+95}), these measurements should shed
light on whether RCW~86 is an older relative of SN~1006, with a Type
Ia progenitor, or a younger version of the Cygnus Loop (a Type II
cavity explosion? See \cite{vkb97}) which we have caught before
significant portions of the shell have begun to develop radiative
shocks.

\acknowledgments

We are grateful to the ever-helpful staff at CTIO who, with help from 
the University of Michigan, have turned the aging Curtis Schmidt back 
into a world class telescope.  We would also like to thank P. Frank 
Winkler for the use of his filters, as well as his helpful comments on 
the paper, and M. J. Kesteven for providing the radio image of the 
remnant.  This work is funded through the generous support of the Dean 
B.\ McLaughlin fellowship at the University of Michigan.

\clearpage

\begin{deluxetable}{lcccccc}
\tablecaption{Journal of Optical Imaging with Curtis Schmidt 
\label{obssum}}

\tablehead{
\colhead{Field} & 
\colhead{RA(2000)} & 
\colhead{Dec(2000)} & 
\colhead{Date (UT)}    & 
\colhead{\ha }    & 
\colhead{\sii }   &  
\colhead{Red }
}

\startdata 
NE    & 14:44:00 & -62:18:00 & 1994 Mar 23  & 5$\times$600s & 3$\times$600s & 4$\times$300s \nl
NW    & 14:40:45 & -62:18:30 & 1994 Mar 24  & 6$\times$600s & 4$\times$600s & 4$\times$300s \nl
SE    & 14:44:00 & -62:36:00 & 1994 Mar 23  & 5$\times$600s & 3$\times$600s & 3$\times$300s \nl
SW    & 14:41:08 & -62:32:00 & 1994 Mar 24  & 4$\times$600s & 3$\times$600s & 4$\times$300s \nl
 \nl
NE    & 14:44:02 & -62:18:15 & 1995 Jan 26  & 3$\times$600s & \ldots & 3$\times$300s \nl
NW    & 14:40:39 & -62:18:34 & 1995 Jan 27  & 3$\times$600s & \ldots & 3$\times$300s \nl
SE    & 14:44:00 & -62:36:15 & 1995 Jan 29  & 4$\times$600s & \ldots & 2$\times$300s \nl
SW    & 14:40:38 & -62:37:24 & 1995 Jan 28  & 3$\times$600s & \ldots & 2$\times$300s \nl
Center& 14:41:35 & -62:32:06 & 1995 Jan 28  & 2$\times$600s & \ldots & 2$\times$200s \nl

\enddata

\end{deluxetable}


\clearpage

\figcaption[Smith.fig1.ps]{The \ha\ image mosaic of the supernova remnant
RCW~86.  The image is 46$\amin$8 square, covering the entire area of
the X-ray and radio shells, with N up and E to the left.  The bright
radiative filaments are immediately apparent.  While some of the
fainter filaments are also noticeable, it is difficult to judge their
extent due to the stellar confusion in this region projected so near
the Galactic plane.\label{hawstars}}


\figcaption[Smith.fig2.ps]{The continuum-subtracted \ha\ mosaic of
RCW~86, showing the wealth of faint Balmer filaments stretching almost
completely around the remnant, as well as faint diffuse \ha\ emission
in the center.  Most of the ``noise'' seen in this image is from the
wings of subtracted stars.  \label{hacsub}}


\figcaption[Smith.fig3.ps]{The final \ha\ mosaic RCW~86, which has been
median filtered to suppress the remains of most of the subtracted
stars.  \label{hamsub}}


\figcaption[Smith.fig4.ps]{The final median-filtered,
continuum-subtracted \sii\ mosaic of RCW~86, clearly showing the
locations of the radiative filaments. \label{siimsub}}


\figcaption[Smith.fig5.ps]{The final \ha\ image mosaic of RCW~86 with
various filament complexes identified for reference.  The labels run
from the center of the remnant outward, so that smaller numbers
correspond to smaller projected distances from the center of the
remnant.  Where needed, suffixes are used to distinguish the
Balmer-dominated filaments (``b'') from the radiative filaments
(``r'').  \label{hawlab}}


\figcaption[Smith.fig6a.ps,Smith.fig6b.ps,Smith.fig6c.ps] {A
comparison of RCW~86 (a) in X-rays, as seen with ROSAT PSPC, (b) in
the optical in \ha\ emission, and (c) in the radio at 847
MHz. \label{3band}}


\end{document}